\documentclass[10pt,twocolumn,conference]{IEEEtran}

\usepackage{graphicx}
\graphicspath{{figures/}}
\usepackage{amsmath}
\usepackage{amssymb}
\usepackage{booktabs}
\usepackage{comment}
\usepackage{enumitem}
\usepackage{xurl}
\usepackage{hyperref}
\usepackage{xcolor}
\usepackage{listings}
\usepackage{algorithm}
\usepackage{algpseudocode}
\usepackage{subcaption}
\usepackage{balance}
\usepackage{fancyhdr}
\usepackage{tikz}
\usetikzlibrary{positioning,arrows,shapes,calc}

\newcommand{\revised}[1]{#1}

\newcommand{\hpcayear}{2027}

\newcommand{\hpcasubmissionnumber}{356}
\title{Making Protocol Effects Explicit: LLM-to-Packet Simulation for Superpod Interconnect Exploration}

\def\hpcacameraready{} 

\newcommand\hpcaauthors{
Yu Zhu\textsuperscript{1}, 
Xinhao Yang\textsuperscript{1}, 
Zhaoshi Li\textsuperscript{2}, 
Baodong Wu\textsuperscript{3}, 
Boxun Li\textsuperscript{3}, 
Wenxue Li\textsuperscript{4}, 
Guohao Dai\textsuperscript{5}, \\
Hongyang Jia\textsuperscript{1}, 
Shuwen Deng\textsuperscript{1}, 
Huazhong Yang\textsuperscript{1}, 
Yuan Xie\textsuperscript{4}, 
Zhenhua Zhu\textsuperscript{1}, and 
Yu Wang\textsuperscript{1}
}
\newcommand\hpcaaffiliation{
\textsuperscript{1}Tsinghua University, 
\textsuperscript{2}MetaX Technology, 
\textsuperscript{3}Infinigence-AI, \\
\textsuperscript{4}The Hong Kong University of Science and Technology, 
\textsuperscript{5}Shanghai Jiaotong University
}
\newcommand\hpcaemail{zhuzhenhua@mail.tsinghua.edu.cn, yu-wang@mail.tsinghua.edu.cn}

\author{
  \ifdefined\hpcacameraready
    \IEEEauthorblockN{\hpcaauthors{}}
      \IEEEauthorblockA{
        \hpcaaffiliation{} \\
        \hpcaemail{}
      }
  \else
    \IEEEauthorblockN{\normalsize{HPCA \hpcayear{} Submission
      \textbf{\#\hpcasubmissionnumber{}}} \\
      \IEEEauthorblockA{
        Confidential Draft \\
        Do NOT Distribute!!
      }
    }
  \fi 
}

\begin{document}
\maketitle


\newcommand{\hpcaheight}{0mm}
\ifdefined\eaopen
\renewcommand{\hpcaheight}{12mm}
\fi

\begin{abstract}
Large language models (LLMs) have become the
foundation of modern AI services. 
The rapid growth of LLM sizes and context lengths has exceeded the memory capacity of a single GPU or NPU, requiring model partitioning across devices.
However, partitioning models across devices introduces costly communication overhead, driving the need for superpods with large-scale, high-bandwidth, low-latency scale-up fabrics.
Exploring superpod designs requires accurate interconnect simulation, as communication latency is jointly determined by link technologies, topologies, and scale-up protocol-stack choices.
Latency-bandwidth models estimate communication latency using a startup term and a transfer term derived from message size and effective bandwidth. But they fail to capture contention and protocol effects, resulting in errors of up to 40\%.
Packet-level simulators like ns-3 model routing, queueing, and packet timing, but require protocol-specific LLM-to-packet implementations. The high simulation time also hinders efficient end-to-end superpod evaluation and design exploration.

This paper presents \emph{ProtoBridge}, an ns-3-based LLM-to-packet simulator for superpod interconnect exploration.  
To make protocol effects explicit, \emph{ProtoBridge} 
introduces two abstraction layers over ns-3 to evaluate different scale-up protocols using a common simulation substrate:
(1) an
operation-to-packet (OTP) layer that captures protocol-specific
packet exchanges and dependencies, and (2) a packet-execution (PEX)
layer that applies shared packet-level mechanisms to the ns-3 event
system. 
\emph{ProtoBridge} achieves average latency error of 17.8\% on H200 NVLink4 and 10.3\% on C500 MetaXLink.  
To enable faster superpod evaluation for further interconnect exploration,
we derive a surrogate model
that combines OTP parameters with analytical models of PEX mechanisms. The
surrogate reduces the mean absolute percentage error (MAPE) from 55.7\% of latency-bandwidth models to 11.8\%, and reduces evaluation time from 2.38$\sim$75.70\,s to 20.5\,$\mu$s per operation.
Using the surrogate-to-simulator evaluation workflow, we explore 38,232 superpod designs across topologies, link technologies, and protocol configurations in one hour.
We use \emph{ProtoBridge} to analyze two emerging superpod designs, scale-up memory pool and co-packaged optics (CPO) interconnects, revealing design insights for future LLM superpod architectures.

\end{abstract}

\section{Introduction}
\label{sec:introduction}

Larger models and longer contexts have pushed LLM inference
beyond the memory capacity of a single GPU or NPU. Serving systems distribute model weights, key-value
(KV) cache, and
activations through tensor, pipeline, context, and expert
parallelism~\cite{megatron_lm,gpipe,deepspeed_moe}. Partitioning relieves GPU
memory pressure but causes collective, point-to-point, and remote-memory
communications that affect inference latency and throughput~\cite{vllm,sglang,distserve,deepspeed_moe}. 
To reduce the communication overhead, 
emerging superpods connect tens to thousands of accelerators through high-bandwidth, low-latency scale-up fabrics.
For example,
NVIDIA GB200 NVL72 connects 72 GPUs in a scale-up network and provide
130\,TB/s of aggregate GPU-to-GPU bandwidth~\cite{nvl72}.

However, superpod interconnect design has not converged. Current products and
prototypes differ in link technology, topology, and scale-up
protocol stacks.
These design choices jointly determine communication latency
and create a large design space that requires accurate and
efficient interconnect evaluation during early-stage architecture exploration.

\begin{figure}[t]
\centering
\includegraphics[width=\linewidth,pagebox=cropbox]{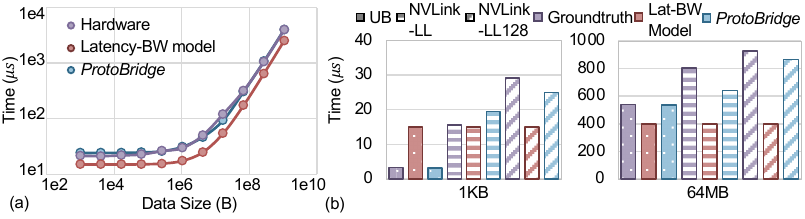}
\caption{AllReduce latencies across (a) message sizes and
(b) NVLink and Unified Bus (UB) protocol configurations.}
\label{fig:intro-motivation}
\end{figure}

Existing approaches are insufficient for evaluating superpod interconnect diversity.
Some system-level simulators estimate communication
latency using \textit{\textbf{latency-bandwidth models}}, where latency is
modeled as a startup cost plus a transfer time determined
by message size and effective bandwidth~\cite{flexflow,proteus}.
However, they ignore packet-level behaviors and
suffer from low accuracy. For example, on an
8-GPU H200 system in Figure~\ref{fig:intro-motivation}(a), the latency-bandwidth model
underestimates 256\,MB AllReduce latency by 38.6\%.
Furthermore, the latency-bandwidth model cannot capture the effects of different scale-up protocol configurations, resulting in large errors across protocols. As shown in Figure~\ref{fig:intro-motivation}(b), different protocols may cause different communication latency for the same data size, which cannot be captured by latency-bandwidth models.
\revised{
As communication latency accounts for a significant fraction of decode time, e.g., 32.4$\%$ for GLM-5.2, accurate communication simulation is required for end-to-end LLM serving evaluations.
}


\begin{figure*}[t]
\centering
\includegraphics[width=\textwidth,pagebox=cropbox]{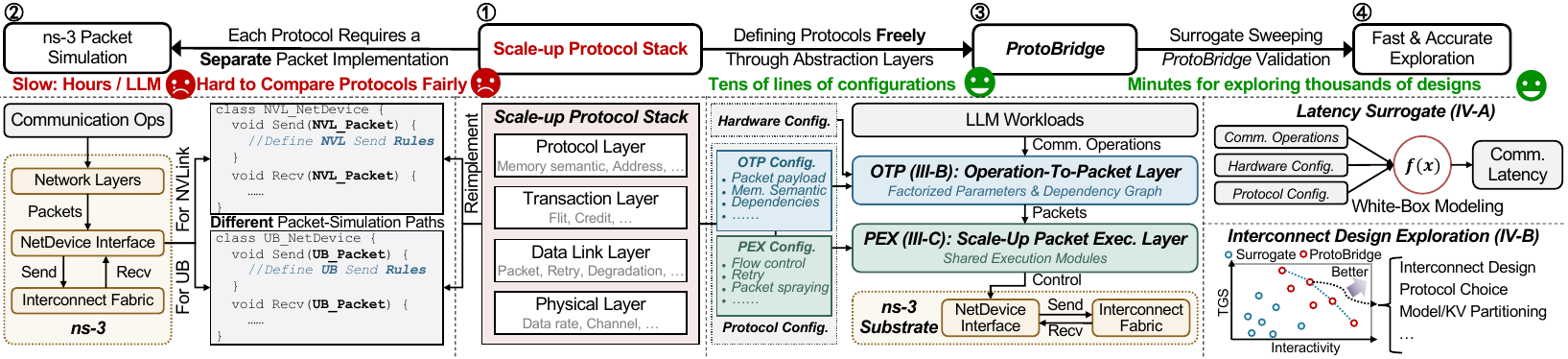}
\caption{\emph{ProtoBridge} architecture and contribution overview. NVL
denotes NVLink;
TGS denotes tokens per GPU per second.}
\label{fig:protobridge-overview}
\end{figure*}

Other \textit{\textbf{packet-level simulators}}, e.g., ns-3~\cite{ns3}, resolve the packet behavior during simulation, but they are ineffective in both end-to-end evaluation and interconnect design exploration for two reasons. First, 
existing packet simulators lack a common abstraction
for scale-up protocols. A new protocol requires
protocol-specific implementations to generate packets, define
completion semantics, and execute packet-level mechanisms.
Consequently, comparing different protocols requires developing
separate simulation paths, making fair evaluation costly and difficult.
Second, packet-level simulation is too slow at LLM serving scale. One
communication operation takes 2.38$\sim$75.70\,s, while one end-to-end LLM
serving configuration invokes 7,869$\sim$24,252 operations. Therefore, packet simulation
would require hours or days to evaluate one serving configuration.
In order to evaluate and find the best superpod interconnect design across numerous topologies and protocols, we
need a simulator with a common protocol abstraction and a
faster estimator that preserves protocol effects
without simulating every packet.

To make the protocol effects explicit and efficiently compare various interconnect designs, we propose \emph{ProtoBridge}, an ns-3-based LLM-to-packet simulator. We introduce two abstraction layers, separating
protocol-specific behavior from the shared ns-3 simulation substrate, as shown in
Figure~\ref{fig:protobridge-overview}. 
The operation-to-packet (OTP) layer first translates each communication step into packet transfers and defines each step's dependencies and completion
rules. The packet-execution (PEX) layer then carries the packet execution modules that can be shared across protocols, and sends the packets to event-based ns-3. 
This separation models different protocols on a common simulation substrate, only requiring the users to change the input configurations rather than reimplementing protocol-specific packet behaviors. \revised{Table~\ref{tab:simulator-positioning} shows comparisons among \emph{ProtoBridge} and other simulators and analytical models.}

\begin{table}[t]
\centering
\caption{\revised{\emph{ProtoBridge} vs. representative simulation and
analytical approaches. $\checkmark$: native support; $\triangle$:
partial support; --: outside the primary scope.}}
\resizebox{\linewidth}{!}{
\label{tab:simulator-positioning}
\begin{tabular}{@{}lcccc@{}}
\toprule
Approach &
LLM Serving &
Op.-to-Pkt. &
Exec. Mechanisms &
Reusable Proto. \\
\midrule
ASTRA-sim~\cite{astra_sim2}
& $\triangle$ & $\checkmark$ & $\triangle$ & -- \\
BookSim~\cite{booksim}
& -- & -- & $\checkmark$ & -- \\
DICE~\cite{dice}
& -- & -- & $\checkmark$ & -- \\
ns-3-UB~\cite{ns3_ub}
& -- & $\triangle$ & $\checkmark$ & -- \\
HARQ~\cite{harq}
& -- & -- & $\triangle$ & -- \\
\emph{\textbf{ProtoBridge}}
& $\checkmark$ & $\checkmark$ & $\checkmark$ & $\checkmark$ \\
\bottomrule
\end{tabular}
}
\end{table}

Furthermore, to support broad superpod evaluation and interconnect exploration,
we derive a protocol-defined latency surrogate model from the 
mechanisms of OTP and PEX layers. The surrogate
extracts the OTP parameters to define the startup latency and effective
bandwidth of the communication operation. It also uses theoretical latency equations from PEX mechanisms to model the effect on communication latency.
Therefore, the surrogate can accurately capture
protocol effects while reducing the modeling cost to several equations.

In this paper, we make the following four contributions.
\begin{itemize}
  \item 
  \revised{
  We separate protocol-defined packet exchanges and completion dependencies from shared packet-execution mechanisms, forming the the operation-to-packet and packet-execution abstraction layers. This abstraction explicitly exposes protocol effects while enabling a unified packet-level simulation substrate.}

  \item We implement \emph{ProtoBridge}, an ns-3-based
    packet-level simulator for LLM superpod interconnects. 
    \emph{ProtoBridge} takes the LLM communication operation and user-defined interconnect (protocol and hardware) as inputs, uses OTP-PEX abstraction layers to generate packets, and simulates the packets using ns-3. \emph{ProtoBridge} achieves mean absolute percentage error (MAPE) of 17.8\%, 10.3\%, and 1.76\% on H200 NVLink4, C500 MetaXLink, and NVL72 across various communication operations, respectively.

  \item 
  \revised{
  We derive a protocol-defined surrogate that composes OTP-defined transfer semantics and  PEX mechanism interactions step by step, capturing their joint effects on communication time. }
  It reduces MAPE from 55.7\% of latency-bandwidth models to 11.8\%, and reduces evaluation time from 2.38\,s to 20.5\,$\mu$s per operation.

  \item We propose a surrogate-to-simulator superpod interconnect evaluation workflow, 
  which efficiently analyzes and explores superpod interconnect choices across link technologies, topologies, and protocols.
\end{itemize}

We apply this surrogate-to-simulator workflow to two emerging
superpod design choices and derive architecture-level insights.
First, we study scale-up memory pools motivated by the growing
memory demand of long-context LLM serving, and analyze the
tradeoff between additional memory capacity and network
contention. Second, we explore co-packaged optics (CPO) for
large-scale superpods and evaluate the impact of link bandwidth,
reach, latency, and error handling on system performance.


\section{Background}
\label{sec:background}

\subsection{LLM Inference Communication}

When model weights, KV cache, and activations exceed the memory of one GPU or
NPU, serving must partition them across devices. Tensor parallelism combines
partial layer results with collectives such as AllReduce, ReduceScatter, and
AllGather~\cite{megatron_lm}. Context parallelism exchanges attention data
across token partitions~\cite{deepseek_v32}. Expert parallelism uses AllToAll
for token dispatch and combination~\cite{deepspeed_moe}, while pipeline stages
exchange activations through point-to-point transfers~\cite{gpipe}. KV cache
migration and prefill/decode disaggregation move KV data between serving
workers~\cite{pd_disagg,distserve,sglang}. A scale-up-attached memory pool
additionally introduces remote-memory reads and writes. These communications
often gate subsequent computation, placing their latency on the serving
critical path.

Communication operations reach the fabric through different logical exchange
patterns. A collective algorithm decomposes a collective into transfers among
the participating devices. For AllReduce, ring, tree, or NVLink SHARP (NVLS)
determines the communicating peers, step order, and bytes sent per
step~\cite{nccl,sharp,nvswitch}. Point-to-point and remote-memory operations
instead use direct transfers or request-response exchanges. The selected
scale-up protocol configuration determines how each exchange is packetized,
controlled, and completed. Together with fabric timing, the resulting packet
traffic and completion condition determine operation latency.
\begin{figure}[t]
\centering
\includegraphics[width=\linewidth,pagebox=cropbox]{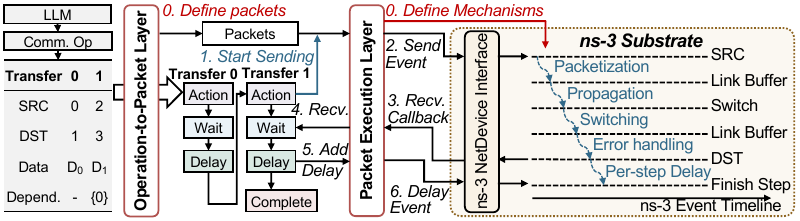}
\caption{\emph{ProtoBridge} execution over the ns-3 event path. SRC and DST
denote source and destination.}
\label{fig:protobridge-execution}
\end{figure}

\subsection{Superpods and Scale-Up Protocols}

A superpod is a large group of GPUs or NPUs connected by a
high-bandwidth, low-latency scale-up fabric. Published systems include NVIDIA
GB200 NVL72 with 72 GPUs, Huawei CloudMatrix 384 with 384 NPUs, and Google TPU
v4 Pod with 4,096 Tensor Processing Unit (TPU)
chips~\cite{nvl72,cloudmatrix384,tpuv4}. Their interconnects differ in system
scale, topology, and protocol stack.
A scale-up protocol configuration combines choices at the transaction,
link, physical, and protocol layers. Transaction layer define
flits, credits, and operation completion. Link control governs packet
framing, ordering, credits, and retransmission, while physical-layer choices
set lane rate, encoding, propagation delay, and forward error correction
(FEC). Protocol layer defines semantics and addresses~\cite{ieee802.3,kp4_fec,ualink_spec}.
These choices jointly determine the packet exchanges produced for an operation
and the time required to execute them.


\section{\emph{ProtoBridge} Structure}
\label{sec:protobridge}

\subsection{Overview}
\label{sec:overview}
\begin{figure}[t]
\centering
\includegraphics[width=\linewidth,pagebox=cropbox]{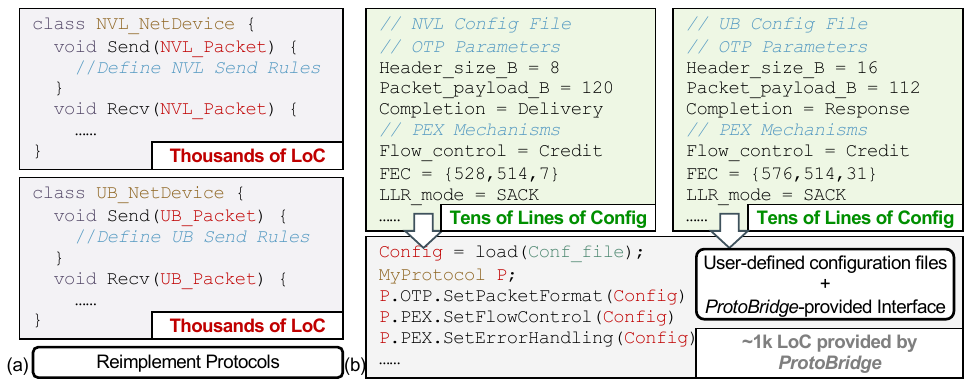}
\caption{(a) Protocol-specific code and (b) the shared \emph{ProtoBridge} interface. LoC represents Lines of Code.}
\label{fig:protocol-interface}
\end{figure}

\emph{ProtoBridge} takes an LLM communication operation, a scale-up protocol
configuration, and an interconnect configuration as input and returns the
operation completion time. To simulate different protocols and interconnections
on a common substrate, \emph{ProtoBridge} builds two abstraction layers on
ns-3, and Figure~\ref{fig:protobridge-execution} summarizes the execution
process. The OTP layer represents the LLM communication operation as a transfer-dependency graph.
When a transfer's dependent nodes are ready, OTP packetizes the transfer and
passes the resulting packets to PEX. The PEX layer applies the configured
flow-control, forwarding, and error-handling rules to each packet, then passes
the packet to ns-3 for transmission over a link. When a packet reaches its
destination, PEX reports its arrival to OTP as an event. OTP uses these events
to determine when the current transfer's complete dependency is satisfied and the next transfer
can begin. 
\revised{As illustrated in Figure~\ref{fig:protocol-interface}, OTP specifies protocol-declared transaction types, communication endpoints, packet formats and sizes, transfer dependencies, and completion conditions, while PEX applies flow control, packet spraying and reordering, link-error injection, FEC, and retransmission. Protocols covered by existing OTP parameters and PEX mechanisms require only configuration changes. New protocol mechanisms extend PEX, and changes below these interfaces such as changing the device model of the switch require modifications to the ns-3 substrate.}

\subsection{Operation-to-Packet Layer}

Protocol configurations can impose different dependencies on the same
transfer. For example, a data \textit{send} completes when the required
bytes reach its destination, while a remote-memory \textit{load} first sends a request
and then completes when the required data return. A common representation must express
various protocols and determine transfer completion from packet arrivals.

We propose the OTP layer to solve this problem by encoding protocol-specific packetization and completion
rules in a \textbf{transfer-dependency graph}.
The graph represents one communication operation, with each transfer
contributing a subgraph. As shown in Algorithm~\ref{alg:protocol-execution}, each transfer subgraph includes four types of nodes: (1) an ACTION node packetizes and
issues a transfer, (2) a WAIT node specifies the data, response, or acknowledgment
required before a data transfer completes, (3) Optional DELAY nodes insert
protocol-defined delays, and (4) a COMPLETE node marks the end of the
operation. Directed edges define the dependencies between nodes. At runtime, a node becomes ready
after all dependent nodes have completed, and as illustrated in Figure~\ref{fig:protobridge-overview}, executing a transfer subgraph requires the follow steps: 
\begin{itemize}
\item The initial ACTION node
passes packets to PEX when the operation begins. 
\item The packets are sent to ns-3 and go through the propagation path.
\item The packet arrivals from ns-3 go through PEX and update the
WAIT node.
\item The DELAY nodes insert fixed latency to the simulation. 
\item When COMPLETE becomes ready, OTP records the current
simulation time as the operation completion time.
\end{itemize}

\begin{algorithm}[t]
\caption{OTP graph execution.}
\label{alg:protocol-execution}
\footnotesize
\begin{algorithmic}[1]
\Require $\mathit{protocol\_config}$, $operation$
\Ensure $\mathit{complete\_timing}$
\State $protocol \gets
    \Call{DefineProtocol}{\mathit{protocol\_config}}$
\State $rules \gets protocol.operation\_rules$
\State $graph \gets \Call{GenerateDependencyGraph}
    {operation, rules}$

\Procedure{OnNodeReady}{$node$}
    \Comment{OTP layer}
    \If{$node.type = \textsc{Action}$}
        \State $format \gets protocol.packetization$
        \State $packets \gets \Call{Packetize}
            {node.transfer, format}$
        \ForAll{$packet \in packets$}
            \State \Call{ExecutePacket}{$packet$}
                \Comment{Handoff to PEX}
        \EndFor
    \ElsIf{$node.type = \textsc{Wait}$}
        \State \Call{Wait}{$node, protocol.completion\_rule$}
    \ElsIf{$node.type = \textsc{Delay}$}
        \State \Call{Delay}{$node$}
    \ElsIf{$node.type = \textsc{Complete}$}
        \State $\mathit{complete\_timing} \gets
            \Call{Complete}{operation}$
        \State \Return $\mathit{complete\_timing}$
    \EndIf
\EndProcedure

\end{algorithmic}
\end{algorithm}

Within OTP, the protocol configuration is specified by the dependecy graph and its node
parameters. Table~\ref{tab:operation-to-packet-parameters} introduces how the
configurations are used in OTP. Packet-transfer and packetization fields configure
ACTION nodes. Completion and local-timing fields configure WAIT and DELAY
nodes. Dependency fields connect transfer sub-graphs into a communication
operation. In this way, various protocol configurations can be simulated
through a common graph executor, and the generated packets are further passed to the PEX layer through the same interface.

\begin{table}[t]
\centering
\caption{Factorized OTP protocol parameters.}
\label{tab:operation-to-packet-parameters}
\footnotesize
\setlength{\tabcolsep}{4pt}
\begin{tabular}{@{}p{0.25\columnwidth}p{0.67\columnwidth}@{}}
\toprule
Parameter group & Parameters \\
\midrule
Packet transfer
  & Transaction and action type, source and destination, packet type,
    effective and response bytes \\
Packetization
  & Header bytes and packet payload \\
Dependencies
  & ACTION, WAIT, DELAY, and COMPLETE nodes with dependency edges \\
Completion
  & Delivery, response, or acknowledgment event \\
Local timing
  & Startup and per-transfer-step processing \\
\bottomrule
\end{tabular}
\end{table}

\subsection{Packet-Execution Layer}
\label{sec:pex}
The scale-up protocol configuration defines packet-execution
mechanisms, including flow control, switch forwarding, forward-error-correction (FEC), and
retransmission. Because multiple mechanisms are applied to the same packet,
they jointly affect the packet transfer time. For
example, an error packet will delay credit return, block later injection, and re-enter the
switch queues when retransmitted. 

\begin{figure}[t]
\centering
\includegraphics[width=\linewidth]{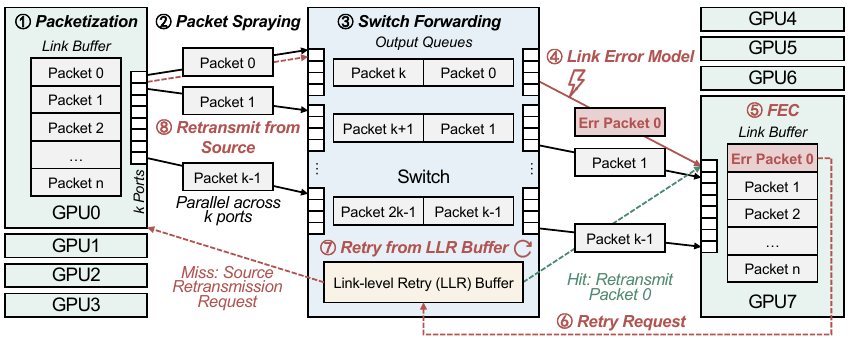}
\caption{Shared scale-up PEX path.}
\label{fig:packet-execution-layer}
\end{figure}

To accurately simulate the packet-execution mechanisms and their interactions, we propose the
PEX layer that composes
the protocol-defined packet mechanisms in one event-driven execution path. PEX allows each protocol
configuration to select the packet-execution mechanisms and their parameters, while different
protocols reuse the shared PEX implementation.
Figure~\ref{fig:packet-execution-layer} illustrates a packet's execution path
through the source device, links, switches, and destination device:
\begin{enumerate}
\item Packet execution begins when PEX receives the packets generated by OTP. The
source device will first check whether the configured flow-control policy permits
each packet to enter the interconnect. 
\item The packets go through  multiple routes to reach the destination if packet spraying is enabled in the configuration.
\item PEX models the switch routing, arbitration, and queuing for the packets.
\item A user-defined link error model is applied to the packets, possibly causing packet errors.
\item At destination, the protocol-defined FEC will try to discover and correct error packets. 
\item When a packet is uncorrectable by the current FEC, it will trigger a retry request.
\item If the required packets hit the link-level retry (LLR) buffer, they will be retransmitted to the destination device.
\item If LLR buffer misses, the packets will be retransmitted by the source device, causing longer latency.
\end{enumerate}

In the PEX path, the protocol-defined packet-execution mechanisms can mainly be categorized into three groups: (1) credit-based flow
control that governs packet injection, (2) switched forwarding that selects the outgoing link
and arbitrates packets waiting for that link, and (3) Link-error handling
that governs FEC and retransmission after a link transfer. We will introduce how PEX implement these mechanisms in the following of this part:

\textbf{Credit-based flow control.}
The PEX layer supports credit-based flow control \revised{with end-to-end credit return}, as used by
NVLink, UB, and UALink~\cite{nvidia_pascal_nvlink,blumrich_nvswitch_credit,ub_base_spec,ualink_spec}.
Each virtual channel has a finite number of credits. Injecting a packet
consumes one credit, but when none remain, the source places the packet in its send
queue. Successful packet delivery sends a credit packet on the reverse path. The
source increments the available-credit count when this packet returns and then
admits queued packets. A delayed delivery therefore postpones credit return
and can block later injection.

\textbf{Switched forwarding.}
At each switch, route selection chooses the outgoing link. Packets waiting for
the same link share an output queue, and switch arbitration selects the next
packet to send. Packets from concurrent operations share these output queues
and are arbitrated together, so concurrent operations can delay each another. The PEX layer
also implements packet spraying.
When equal-cost paths are available, packet spraying distributes a
transfer across them. A destination reorder buffer restores packet order
before reporting the transfer's delivery.

\textbf{Link-error handling.}
Our link model combines FEC with optional link-level retry
(LLR), which are widely used by high-speed scale-up links~\cite{kp4_fec,ualink_spec}.
Each link samples errors from its configured pre-FEC BER and applies
the selected FEC code. In the FEC$(N,K,T)$ model, a codeword contains $N$
symbols, of which $K$ carry data, and corrects up to
$T=\lfloor(N-K)/2\rfloor$ symbol errors~\cite{rs_codes,kp4_fec}. Coding
expands the transmitted bytes and adds encoder and decoder latency defined in the configuration file. After FEC
decoding, PEX delivers a packet when the errors can be corrected.

When the errors exceed the FEC correction capability, PEX marks the packet as uncorrectable
and returns a negative acknowledgment (NACK) back to the sender. Selective acknowledgment (SACK)~\cite{sack_tcp} and Go-Back-N (GBN)~\cite{gbn_original} are two typical retransmission mechanisms used in superpods. SACK resends the failed
packet. GBN also resends the later
outstanding packets. The sender retains copies of unacknowledged packets in a
finite retransmission buffer. If the required copy has been evicted, PEX
reloads its payload from the source memory (e.g., HBM) before reinjection. The
reload adds the source-memory access latency according to user's configuration and causes the time to re-transmit the packet.

\revised{
\textbf{Extensibility and implementation effort.} Protocol extensions follow two cases: a new protocol can reuse existing OTP parameters and PEX mechanisms through configuration, whereas a previously unsupported PEX mechanism requires a PEX extension. We demonstrate the first case by implementing Scale-Up Ethernet (SUE) from its public specification with a 50-line configuration file. SUE reuses the existing OTP interface and PEX mechanisms for credit flow control, Reed–Solomon FEC, and selective retransmission. For the second case, we implement window-based flow control by adding per-endpoint window-size and outstanding-packet counters to the PEX send gate, dispatching packet admission according to the selected flow-control policy, decrementing the outstanding count upon delivery feedback, and reusing the existing send queue to inject pending packets after a window slot is released. It adds 42 lines of code to PEX.
Compared with implementing a protocol directly in the original ns-3, \emph{ProtoBridge} confines protocol extensions to the boundary defined above but substantially reduces development effort. For example, SUE-Sim adds $\sim$9k lines of code to ns-3 to implement SUE protocol.}

\section{Surrogate-to-Simulator Evaluation Flow}
\label{sec:surrogate}
\emph{ProtoBridge} provides accurate packet-level simulation for a given superpod design. But emerging superpods introduce a large and diverse design space across topologies, link technologies, and protocols, which cannot be efficiently explored with packet-level simulation alone. For example, the CPO case study in Section~\ref{sec:case-cpo} considers 531 interconnect designs and 72 error-handling configurations, i.e., 38,232 design points. Exhaustively evaluating these candidates with packet-level simulation is prohibitively expensive. Therefore, we develop a surrogate-to-simulator evaluation flow that uses a fast analytical surrogate to explore the design space and applies \emph{ProtoBridge} for accurate validation of promising designs.


\subsection{Protocol-Defined Latency Surrogate}
\label{sec:surrogate-theory}

Packet simulation spends most of its runtime on scheduling individual packet
events. To enable diverse superpods evaluation and exploration, we propose a surrogate model that simplifies the events while keeping the two quantities
exposed by \emph{ProtoBridge}: the transfer parameters defined by OTP and the
latency effects of the mechanisms selected in PEX. Analytical equations then
compute operation latency directly from the protocol configuration without executing every packet event. \revised{The
surrogate divides operation latency into \textbf{protocol startup}, \textbf{fixed delay} on the
critical path of each transfer, \textbf{serialized transfer time}, and \textbf{recovery overhead}.}
And for the transfer time, flow-control stalls, FEC wire overhead, and contention change the effective
throughput of the transfer. This decomposition
gives:
\begin{equation}
\label{eq:surrogate-latency}
\begin{aligned}
\widehat{L}={}&S_p+k_a T_{\mathrm{fixed}} +k_a T_{\mathrm{serial}}\Phi_{\mathrm{credit}}
  \Phi_{\mathrm{fec}}\Phi_{\mathrm{share}} \\
&+T_{\mathrm{retx}}+T_{\mathrm{reload}}+T_{\mathrm{fec,lat}}.
\end{aligned}
\end{equation}
Here, $\widehat{L}$ is the estimated operation latency, $S_p$ is protocol startup
latency, $k_a$ is the number of collective-algorithm transfer steps,
$T_{\mathrm{fixed}}$ is the propagation or switch-pipeline delay per transfer, and
$T_{\mathrm{serial}}$ is the time to transmit the transfer's data. The factors
$\Phi_{\mathrm{credit}}$, $\Phi_{\mathrm{fec}}$, and $\Phi_{\mathrm{share}}$
represent credit pressure, FEC wire expansion, and shared-resource contention,
respectively.
$T_{\mathrm{retx}}$ is the expected retransmission delay,
$T_{\mathrm{reload}}$ is the delay for packets absent from the retry buffer, and
$T_{\mathrm{fec,lat}}$ is the FEC encoding and decoding latency. Each $\Phi$-factor
equals one when the corresponding mechanism adds no latency
and exceeds one when that mechanism reduces the effective throughput of the
transfer. In the following subsection, we will introduce the modeling for each part of the surrogate.

\textbf{Protocol steps and packet transfer.}
Computing Equation~\ref{eq:surrogate-latency} first requires the startup
latency $S_p$, effective bandwidth $BW_{\mathrm{eff}}$, number of transfer steps $k_a$, and bytes sent in each transfer step $B_{\mathrm{step}}$. The collective
schedule determines $k_a$ and $B_{\mathrm{step}}$ from the
operation, participant device count, and algorithm. E.g., an 8\,KB 8-GPU Ring-AllReduce requires 14 steps with $B_{\mathrm{step}}$ equals 1\,KB. The time to
transmit the data in one transfer step is:
\begin{equation}
\label{eq:step-latency}
T_{\mathrm{serial}} =
\frac{B_{\mathrm{step}}}{BW_{\mathrm{eff}}}.
\end{equation}
Both $BW_{\mathrm{eff}}$ and $T_{\mathrm{fixed}}$ come from either a
calibrated platform profile or the hardware
specification. However, using the hardware specification may be less accurate than a calibrated profile.

\textbf{Credit pressure.}
Credit flow control limits the amount of data that a sender can keep in
flight. With $C$ credits and $B_{\mathrm{packet}}$ bytes per credit, the
sender can inject at most $CB_{\mathrm{packet}}$ bytes before a credit
returns. Let $C_{\mathrm{bdp}}$ denote the bandwidth-delay product expressed
in credits. Keeping a link with bandwidth $BW_{\mathrm{link}}$ busy during the
credit-return time $T_{\mathrm{creditRTT}}$ requires:
\begin{equation}
\label{eq:credit-bdp}
C_{\mathrm{bdp}} =
\frac{BW_{\mathrm{link}}T_{\mathrm{creditRTT}}}
     {B_{\mathrm{packet}}}
\end{equation}
credits. \revised{Here we use experience parameters (i.e., 0.4$\mu s$ link latency and 1.1$\mu s$ switch latency) to calculate $T_{\mathrm{creditRTT}}$.} When $C<C_{\mathrm{bdp}}$, the sender exhausts its credits and leaves
the link idle while waiting for their return. The resulting
serialization-latency factor is:
\begin{equation}
\label{eq:credit-factor}
\Phi_{\mathrm{credit}} =
\max\left(1,\frac{C_{\mathrm{bdp}}}{C}\right).
\end{equation}
The ratio $C/C_{\mathrm{bdp}}$ is the fraction of link bandwidth that the available
credit window can sustain, and a factor of one indicates that the credit window is large enough to sustain
the configured link bandwidth.

\textbf{FEC and link-level retransmission.}
FEC changes both the transmitted bytes and the probability that a packet
requires retransmission. We model FEC using a Reed-Solomon
(RS)$(N_{\mathrm{fec}},K_{\mathrm{fec}},T)$ code~\cite{rs_codes}.
$N_{\mathrm{fec}}$ is the number of symbols per codeword,
$K_{\mathrm{fec}}$ is the number of data symbols, and $T$ is the number of
symbol errors the code can correct. If a fraction $f_l$ of the transmitted
bytes traverses the links protected by the selected code, the
FEC-expansion factor is:
\begin{equation}
\label{eq:fec-expansion}
\Phi_{\mathrm{fec}} =
1 + f_l\left(\frac{N_{\mathrm{fec}}}{K_{\mathrm{fec}}}-1\right).
\end{equation}
Besides $(N_{\mathrm{fec}},K_{\mathrm{fec}},T)$, $T_{\mathrm{fec,lat}}$ should also be configured to add the FEC encoding and decoding
pipeline latency.

\emph{ProtoBridge} and the surrogate read the FEC parameters
$(N_{\mathrm{fec}},K_{\mathrm{fec}},T)$ and $T_{\mathrm{fec,lat}}$ from the same protocol
configuration. The surrogate uses the FEC parameters with the link BER to compute the probability
that FEC cannot correct a packet. For pre-FEC BER $b$, a 10-bit
symbol is erroneous with probability $p_s=1-(1-b)^{10}$. Let
$X\sim\mathrm{Binomial}(N_{\mathrm{fec}},p_s)$ be the number of erroneous
symbols in one codeword. The codeword is uncorrectable with probability
$p_{\mathrm{cw}}=\Pr[X>T]$. A $B_{\mathrm{packet}}$-byte packet contains
$Q=\lceil 8B_{\mathrm{packet}}/(10K_{\mathrm{fec}})\rceil$ codewords.
Assuming independent codeword errors, the packet is uncorrectable with
probability:
\begin{equation}
\label{eq:post-fec-packet-error}
p_u = 1-(1-p_{\mathrm{cw}})^Q.
\end{equation}
A transfer step completes only after all packets arrive,
so its retransmission delay is determined by the packet that needs the most
attempts. Let $A_j$ be the number of attempts for packet $j$ and
$A_{\max}=\max_j A_j$. Because $\Pr[A_j>r]=p_u^r$ and $A_j$s are independent variables, 
$\Pr[A_{\max}>r]=1-(1-p_u^r)^P$. The tail-sum identity then gives:
\begin{equation}
\label{eq:retransmission-d}
\mathbb{E}[A_{\max}]
=\sum_{r=0}^{\infty}\left[1-\left(1-p_u^r\right)^P\right].
\end{equation}
SACK re-sends only the failed packets and therefore adds the following
completion-path delay:
\revised{
\begin{equation}
\label{eq:sack-delay}
T_{\mathrm{retx,SACK}}
=k_a\frac{T_{\mathrm{serial}}}{P}\left(\mathbb{E}[A_{\max}]-1\right).
\end{equation}
}
GBN retransmits from the failed packet through the end of the issued
window. The model caps this window at
$W=\min(C,C_{\mathrm{bdp}},P)$, where $C$ is the available credit count,
$C_{\mathrm{bdp}}$ is the credit window needed to fill the path, and $P$ is
the number of packets in the step. Assuming that the failed position is
uniformly distributed within the window, the expected retransmitted suffix
contains $(W+1)/2$ packets.

Retry-buffer capacity determines whether the required packet copies remain in
the buffer at the sender's port. \revised{Let $B_{\mathrm{buf}}$ be the buffer capacity and the buffer totally holds
$C_{\mathrm{buf}}=\lfloor B_{\mathrm{buf}}/B_{\mathrm{packet}}\rfloor$ packet copies. We use $R$ to represent the number of packet copies the schedule requires the sender to retain, e.g., $R = P$ for a single active transfer and $R = 2P$ for a tree step with two active transfers.} Then the fraction of absent packets from the port's buffer is:
\begin{equation}
\label{eq:retry-buffer-miss}
m=\max\left(0,1-\frac{C_{\mathrm{buf}}}{R}\right).
\end{equation}
Let $E_{\mathrm{retry}}$ be the expected number of packets that must be re-sent.
The surrogate computes $T_{\mathrm{reload}}$ from
$mE_{\mathrm{retry}}$ source reads. Each read costs
$L_{\mathrm{src}}+B_{\mathrm{packet}}/BW_{\mathrm{src}}$, and the source memory
latency $L_{\mathrm{src}}$ and bandwidth $BW_{\mathrm{src}}$ come from the
selected memory profile.

\textbf{Shared-resource contention.}
Concurrent packets contend when their routes use the same link or switch
output. Prior flow-level simulation models this contention through bottleneck
rate allocation~\cite{nguyen_maxmin}. \revised{Following the max-min rate allocation in~\cite{nguyen_maxmin}, our surrogate calculate $\rho_{\mathrm{eff}}$, which is the fraction of bottleneck bandwidth that the competing traffic occupies.} 
The evaluated transfer receives the remaining bandwidth,
$BW_{\mathrm{eff}}(1-\rho_{\mathrm{eff}})$, giving
$\Phi_{\mathrm{share}}=1/(1-\rho_{\mathrm{eff}})$. We have
$\rho_{\mathrm{eff}}=0$ and $\Phi_{\mathrm{share}}=1$ if there is no contention.

\begin{figure}[t]
\centering
\includegraphics[width=\linewidth]{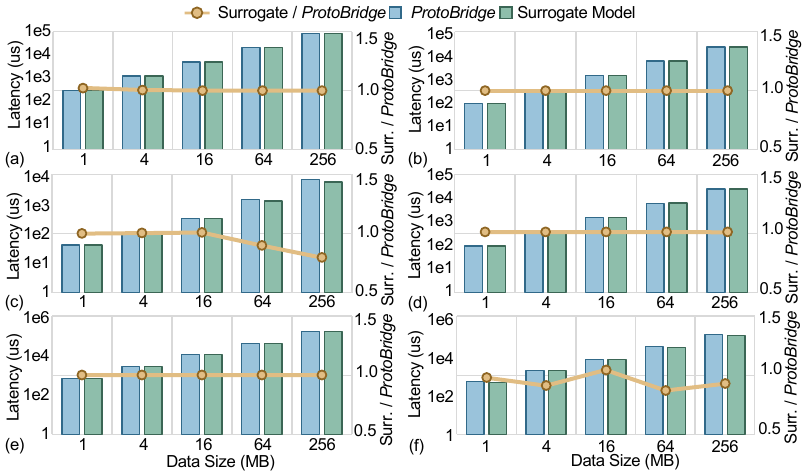}
\caption{\revised{Validations of (a) credit pressure, (b) FEC coding,
(c) high-BER retransmission, (d) output contention, (e) credit and contention, and (f) credit and retry.}}
\label{fig:mechanism-formulas}
\end{figure}

\revised{We evaluate the error of each individual PEX mechanism and the joint effects of credit with contention and credit with retry of the surrogate in Figure~\ref{fig:mechanism-formulas}.} 
\revised{
Surrogate MAPE is 0.6\% for credit pressure, 0.0\% for FEC expansion, 5.2\%
for SACK retransmission at pre-FEC BER $10^{-3}$, 1.1\% for output
contention, 0.01\% for credit with contention, and 7.2\% for credit with retry.
}
The FEC experiment isolates deterministic wire expansion, and the formula
uses the same RS codeword rounding as \emph{ProtoBridge}, so they have exactly the same wire efficiencies.

\begin{figure}[t]
\centering
\includegraphics[width=\linewidth]{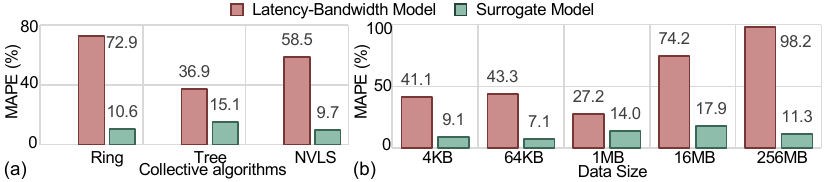}
\caption{Modeling error with different: (a) AllReduce algorithms and (b) transfer sizes.}
\label{fig:surrogate-breakdown}
\end{figure}

\revised{In Figure~\ref{fig:surrogate-breakdown}, we evaluate latency-bandwidth model and our surrogate on various collective algorithms, GPU numbers, credit limits, and BER.} By expressing protocol mechanisms as analytical latency terms, the surrogate
reduces the latency-bandwidth baseline's MAPE from
36.9\%$\sim$72.9\% to 9.7\%$\sim$15.1\% across collective algorithms and
from 27.2\%$\sim$98.2\% to 7.1\%$\sim$17.9\% across transfer sizes. \revised{For small transfers, communication time is dominated by the fixed latency term. As the transfer size increases, the effects of flow control, link errors, and other packet-execution mechanisms accumulate, but the latency–bandwidth model does not capture them. Its modeling error therefore tends to increase with transfer size.} Meanwhile, compared with
\emph{ProtoBridge}, the surrogate reduces per-operation evaluation time from
2.38$\sim$75.70\,s to 20.5\,$\mu$s.

\subsection{Surrogate-to-Simulator Interconnect Evaluation}
\label{sec:system}
Evaluating an emerging superpod architecture requires considering interconnect designs including network topologies, switch buffer sizes, and link technologies, under different protocol choices that include flow control, switch routing, and error handling configurations.
The large interconnect design space makes packet simulation for every design
prohibitively time-consuming. We propose a surrogate-to-simulator evaluation flow that first uses the surrogate to
evaluate the full interconnect design space and select the best serving points for each
design. Then for each LLM workload, we use \emph{ProtoBridge} to evaluate the union of
the top 15\% of designs ranked by surrogate-predicted interactivity and tokens-per-GPU-per-second (TGS) to rank the designs more accurately. 

\begin{figure}[t]
\centering
\includegraphics[width=\linewidth,pagebox=cropbox]{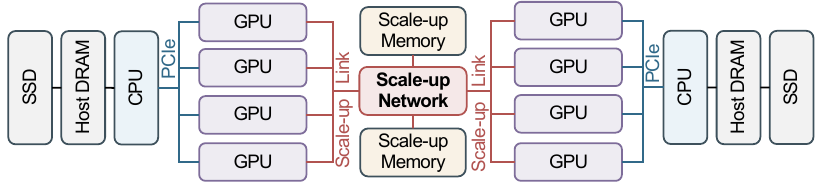}
\caption{Memory paths in the serving evaluation.}
\label{fig:memory-serving-path}
\end{figure}

Communication latency alone cannot rank superpod designs because compute,
memory, and stage overlap determine whether a communication delay
becomes the LLM serving critical path. 
Therefore, we use ASTRA-sim's system-level model to schedule the compute, memory, and communication operations~\cite{astra_sim2}, with the consideration of their dependencies and overlaps.
For each evaluation, we first verify memory capacity and
place model weights and KV data across GPU high-bandwidth memory (HBM) and
the selected host or pooled-memory tier. In this paper, we model GPU computation and HBM access using
a roofline model~\cite{roofline}, while other models with higher accuracies such as Accel-sim~\cite{accel} or GPGPU-sim~\cite{gpgpu} are also supported. If HBMs cannot hold all the data, host DRAM or SSDs will be used. Some existing work~\cite{scalepool} also provides a scale-up
memory pool to prevent high-latency host memory accesses. The memory pool provides extra capacity but adds remote-memory traffic to the
interconnect, as shown in Figure~\ref{fig:memory-serving-path}. The surrogate
and \emph{ProtoBridge} both supply the latency of each communication and
remote-memory operation. 

The prefill and decode time determines interactivity and TGS, which are widely used to evaluate a serving system.
For a request with $P$ input and $O$ output tokens, time to first token
(TTFT), denoted $T_{\mathrm{TTFT}}$, ends when the first output token is
produced, and $T_{\mathrm{decode}}$ covers the remaining $O\!-\!1$ output tokens.
Time per output token (TPOT) and interactivity are:
\begin{equation}
\mathrm{TPOT}=\frac{T_{\mathrm{decode}}}{O-1},
\qquad
\mathcal{I}=\frac{1}{\mathrm{TPOT}}
=\frac{O-1}{T_{\mathrm{decode}}}.
\label{eq:interactivity}
\end{equation}
Interactivity is reported in generated tokens per second for one user. With
$R$ active replicas, a synchronous batch of $B$ requests per replica, and
$G$ GPUs, we define TGS, measured in tokens per GPU per second, as:
\begin{equation}
\mathrm{TGS}=
\frac{RB(P+O)}
{G\left(T_{\mathrm{TTFT}}+T_{\mathrm{decode}}\right)}.
\label{eq:tgs}
\end{equation}

\section{Evaluation}
\label{sec:evaluation}

\subsection{Evaluation Setup}
\label{sec:evaluation-setup}


\textbf{Simulation accuracy setup.}
Table~\ref{tab:evaluation-platforms} summarizes setup for simulation accuracy evaluation.
We compare \emph{ProtoBridge}'s results on H200 and C500 with real hardware profiling, and on NVL72 with open-source data~\cite{coreweave_nvl72_nccl}.
H200 connects eight GPUs with an NVSwitch, and NVL72 uses an 18-plane full-bisection switch topology~\cite{nvl72}.
C500 connects four GPUs through a direct MetaXLink full mesh, and eight C500s are connected through Peripheral Component
Interconnect Express (PCIe) 5.0 host path. 
For UB, the official ns-3-UB simulator~\cite{ns3_ub} supplies the UB results, adding a third scale-up
protocol family and an eight-by-eight two-dimensional full-mesh topology. \revised{Simulators run on Intel Xeon Platinum 8358 CPU at 2.60 GHz, with single-thread for surrogate and eight workers in parallel for \emph{ProtoBridge}.}

\begin{table}[t]
\caption{Simulation accuracy evaluation setup.}
\label{tab:evaluation-platforms}
\footnotesize
\centering
\setlength{\tabcolsep}{2.5pt}
\renewcommand{\arraystretch}{1.05}
\begin{tabular}{@{}p{0.13\linewidth}p{0.48\linewidth}p{0.31\linewidth}@{}}
\toprule
Reference & Fabric configuration & Workload \\
\midrule
H200
& 8 GPUs; NVLink4/NVSwitch; 18 lanes per GPU; 200\,Gb/s per GPU-plane link
& NCCL collectives; 1\,KB$\sim$1\,GB \\
C500
& 4 GPUs: direct full mesh; 8 GPUs: two four-GPU domains and an aggregate PCIe 5.0 host path
& MCCL collectives; 1\,KB$\sim$2\,GB \\
NVL72
& 72 GPUs; 18 NVSwitch planes; 400\,Gb/s per GPU-plane link
& NCCL AllReduce; 512\,MB$\sim$8\,GB
\\
ns-3-UB
& 64 NPUs; 8$\times$8 two-dimensional full mesh; 448\,Gb/s
& Collective operations from 1\,KB$\sim$1\,GB \\
\bottomrule
\end{tabular}
\end{table}

\textbf{Packet-simulation baselines.}
We compare \emph{ProtoBridge} with the analytical, ns-3, and HTSim network
backends of ASTRA-sim~\cite{astra_sim,astra_sim2}.
The released ASTRA-sim with HTSim backend parses only fat-tree topologies. Therefore, we add a graph
loader to HTSim for the C500, UB, and NVL72 physical graphs without changing ASTRA-sim's
collective generation. \revised{The peak bandwidth and startup latency of \emph{ProtoBridge} are calibrated from H200 and C500 measured data, and the baselines receive the same data. These two parameters absorb NCCL software overhead.}

\textbf{Surrogate baselines.}
We evaluate the surrogate on AllReduce configurations with 8 and 16 H200 GPUs.
The configurations cover ring, tree, and NVLS algorithms, with message sizes from
4\,KB to 256\,MB. 
We compare the surrogate with a latency-bandwidth model and three machine learning models: random forest (RF),
XGBoost (XGB), and a multilayer perceptron (MLP). The machine learning models receive the
message size, GPU count, algorithm, credit limit, BER, FEC setting,
and retransmission setting. Random
forest uses 200 trees, XGB uses 300 depth-6 trees, and the MLP uses hidden layers
with 64 and 32 neurons. 

\subsection{Simulation Accuracy}
\label{sec:simulation-accuracy}

\begin{figure*}[t]
\centering
\includegraphics[width=\textwidth,pagebox=cropbox]{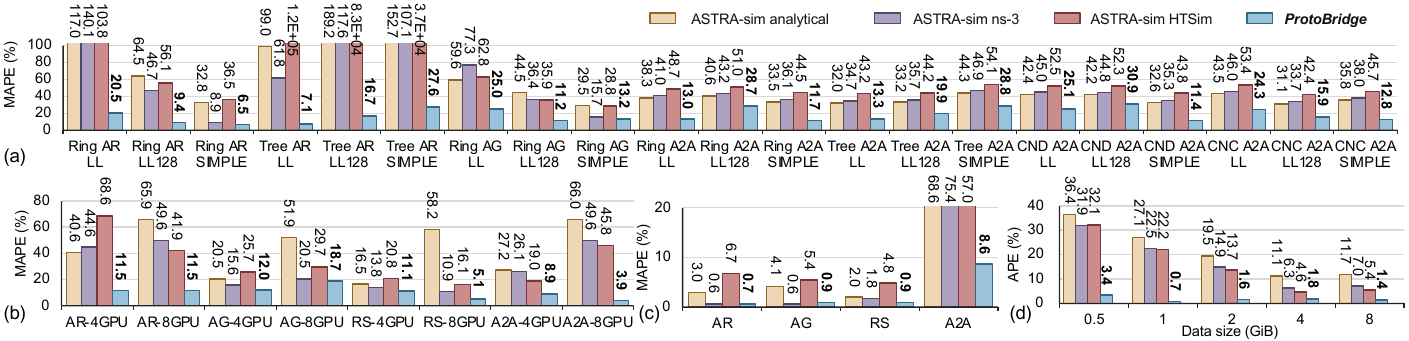}
\caption{Accuracy against H200, C500, NVL72, and ns-3-UB references. AR, AG,
and A2A denote AllReduce, AllGather, and AllToAll; CND and CNC denote
CollNetDirect and CollNetChain.}
\label{fig:simulation-accuracy}
\vspace{-15pt}
\end{figure*}

\begin{figure}[t]
\centering
\includegraphics[width=\linewidth,pagebox=cropbox]{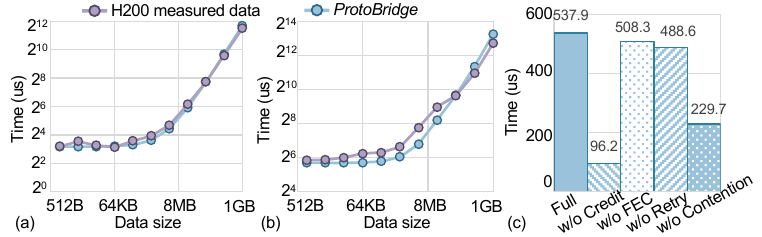}
\caption{\revised{H200 simulation error breakdown by data size for (a) CND A2A SIMPLE and (b) Tree AR SIMPLE. (c) \emph{ProtoBridge} ablation.}}
\label{fig:simulation-ablation}
\end{figure}

\textbf{Overall accuracy.}
We first test whether \emph{ProtoBridge} can accurately simulate different protocol families
and topologies.
Figure~\ref{fig:simulation-accuracy} compares 34 configurations and
shows the MAPE across various data sizes. 
\emph{ProtoBridge}'s mean MAPE is 17.76\% on H200, 10.33\% on C500, 1.76\%
on NVL72, and 2.77\% on UB, while the best ASTRA-sim backend reaches 52.01\%,
28.85\%, 15.62\%, and 18.49\%, respectively.
\revised{Figure~\ref{fig:simulation-ablation}(a) and (b) uses two examples to show the source of error. Large errors occur at medium data sizes, which are caused by the NCCL software overhead not modeled at the protocol layer. At 16KB to 1MB data sizes, software overhead such as chunk DMA, fence, and nProc barrier sync is large relative to transfer time, giving higher MAPE.}


\textbf{Effects of scale-up protocols.}
Existing simulators cannot accurately model the effects of scale-up protocols.
As shown in Figure~\ref{fig:simulation-accuracy}(c), 
all three backends of ASTRA-sim can well model the ring algorithm for AllReduce, AllGather, and
ReduceScatter with little contention. However,
UB AllToAll instead traverses intermediate NPUs in the 2D-full-mesh fabric, causing severe contentions.
The analytical backend replace the multi-hop paths with a fully connected
approximation and incur 68.59\% error. HTSim and ns-3 backends model the 2D-full-mesh, but use the DCQCN flow control~\cite{dcqcn}, which
differs from UB credit flow control
and retains 56.96\% and 75.35\% errors, respectively.
\emph{ProtoBridge} models credit-based flow control and contention in the shared PEX
layer and reaches 8.65\% MAPE in UB AllToAll.
\revised{Figure~\ref{fig:simulation-ablation}(c) provides ablation across PEX mechanisms under 4MB tree AllReduce with $10^{-8}$ BER and 50GB/s contended traffic. Without the simulation of PEX mechanisms, the error can be up to 82\%.}

\subsection{Surrogate Fidelity}
\label{sec:surrogate-evaluation}

Superpod design exploration depends on both latency error and candidate ordering. We
therefore evaluate the surrogate's absolute error, generalization,
ranking accuracy, and the number of candidates that require packet simulation
to find the lowest-latency design.

\begin{figure}[t]
\centering
\includegraphics[width=\linewidth,pagebox=cropbox]{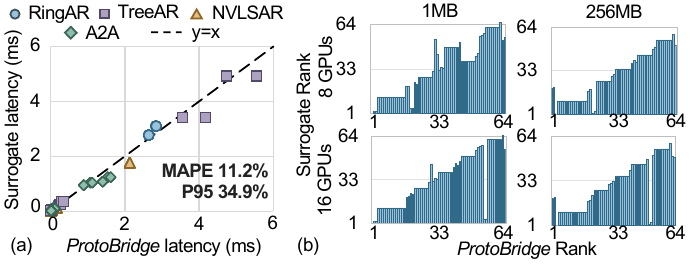}
\caption{\revised{(a) Surrogate latency prediction and (b) design ranking. P95 denotes the
95th-percentile error.}}
\label{fig:surrogate-prediction-ranking}
\vspace{-10pt}
\end{figure}

\begin{figure}[t]
\centering
\includegraphics[width=\linewidth,pagebox=cropbox]{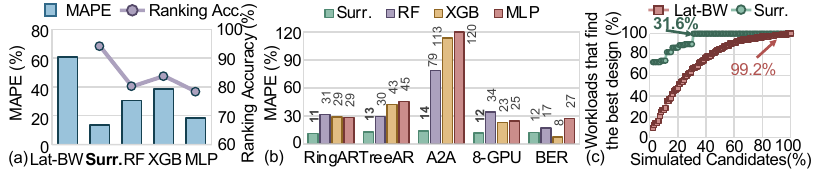}
\caption{\revised{Surrogate fidelity and design selection.}}
\label{fig:mechanism-surrogate}
\end{figure}

\textbf{Prediction and ranking accuracy.}
\revised{The evaluation contains 860 configurations spanning 8 and 16 GPUs,
ring, tree, and NVLS AllReduce and CND AlltoAll algorithms, and message sizes of 4\,KB, 64\,KB, 1\,MB,
16\,MB, and 256\,MB.} The configurations use credit limits of 32 or 128,
pre-FEC BERs from 0 to $10^{-4}$, and
enabled or disabled FEC and link-level retransmission. The machine learning
baselines RF, XGB, and MLP use five-fold cross-validation over distinct
configurations. 
Figure~\ref{fig:mechanism-surrogate}(a) reports both prediction error and
ranking accuracy. \revised{The surrogate reaches 13.4\% MAPE, compared with 60.6\% for
the latency-bandwidth model, while RF, XGB, and MLP reach 30.7\%, 38.4\%, and
28.3\%, respectively. The surrogate also achieves the highest ranking accuracy, while the latency-bandwidth model cannot distinguish different protocol mechanism choices.}
Figure~\ref{fig:surrogate-prediction-ranking}(a) compares every surrogate
prediction with its \emph{ProtoBridge} latency, and Figure~\ref{fig:surrogate-prediction-ranking}(b) shows
the rank assignments at 1\,MB and 256\,MB with each bar represents one
configuration. \revised{Across the complete
evaluation, 95\% of the surrogate-predicted results have an error of less than 34.9\%, and the surrogate orders 87.8\% of the candidate pairs correctly.} Explicit protocol and packet-mechanism terms in our
surrogate model therefore improve both latency prediction and design ranking
over the latency-bandwidth model and the three learned baselines.

Figure~\ref{fig:mechanism-surrogate}(b) evaluates groups excluded when
training the machine learning baselines. Each test excludes one collective algorithm, GPU
count, or error regime.
Withholding one group of training data significantly effects machine learning baselines' accuracies, while our surrogate model keeps the accuracy steady.

\textbf{Finding the lowest-latency design.}
We next measure how many packet simulations are required to identify the
lowest-latency design.
We use the same configurations as the prediction and ranking evaluation. For each GPU-count and message-size pair, the surrogate ranks the candidate protocol configurations, and \emph{ProtoBridge}
simulates them in that order. At each point, we check whether the simulated
candidates include the lowest-latency candidate from the complete set. 
\revised{Figure~\ref{fig:mechanism-surrogate}(c) shows that the surrogate finds the
lowest-latency candidate after simulating 31.6\% of the candidates on average.
The latency-bandwidth model requires 99.2\%.}

\begin{figure}[t]
\centering
\includegraphics[width=\linewidth,pagebox=cropbox]{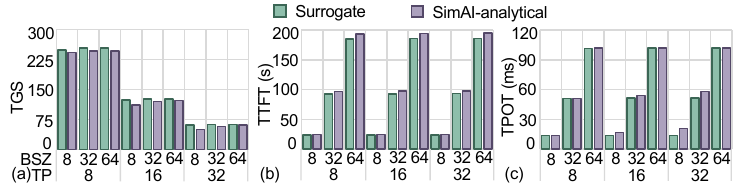}
\caption{\revised{End-to-end comparison on (a) TGS, (b) TTFT, and (c) TPOT between SimAI and surrogate.}}
\label{fig:simai}
\end{figure}

\revised{
\textbf{End-to-end validation.} We provide end-to-end comparison on DeepSeek-V3 with various parallelism and batch sizes between the surrogate model and SimAI~\cite{simai} in Figure~\ref{fig:simai}. SimAI uses analytical communication backend, combining latency-bandwidth model with calibrated bandwidth efficiency on different data sizes. Our surrogate achieves 4.5$\%\sim$7.9$\%$ end-to-end MAPE on TGS, TTFT, and TPOT.
Besides, Sim-AI requires $\sim$16\,ms runtime per operation, while our surrogate only requires $\sim$21\,$\mu$s.}

\section{Case Studies}
\label{sec:case-studies}

We apply \emph{ProtoBridge} and the surrogate to two designs motivated by
emerging serving requirements: (1) Long-context LLMs can exceed aggregate GPU
HBM, motivating memory pools attached to the scale-up fabric, but pool accesses
contend with LLM communication; (2) Larger superpods demand more
interconnection bandwidth and longer-reach links, motivating CPO, but optical
links add conversion and error-handling overhead.

\subsection{Case Study I: Scale-Up Memory Pools}
\label{sec:case-memory-pool}

\subsubsection{Setup}

We compare the systems with and without scale-up-attached memory pools to determine
which designs retain their serving benefit under shared-fabric contention.

\textbf{Design space.}
We evaluate DeepSeek-V3, GLM-5.2, and Qwen3-235B on 128 H200 GPUs, each with
141\,GB of HBM. Each request contains one million context, 8192 input tokens, and generates
1,024 output tokens. The per-replica batch sizes are from 1 to 512. We evaluate colocated and prefill/decode-disaggregated
execution~\cite{pd_disagg,distserve}. Disaggregated strategies partition 128 GPUs between
prefill and decode, varying the allocation from $P/D=8/120$ to $120/8$ in
eight-GPU increments. Replicated variants assign
16, 32, or 64 GPUs to each prefill/decode replica pair. Across the three LLMs,
the partition optimizer considers 2,353 legal combinations of parallelism strategies. For each
design and batch size, we select the partition with the highest TGS at each
interactivity requirement. 
Expert-parallel configurations use DeepEP normal mode during
prefill and low-latency mode during decode~\cite{deepep}.

The interconnect space begins with 23 no-pool configurations from ten topology
families. The
direct topology families include ring, hypercube, and two-dimensional
mesh~\cite{dally_interconnection}.
The switched topology families are multiplane~\cite{gherghescu_multiplane},
NVL-style~\cite{nvl72}, leaf-spine and three-level
Clos~\cite{alfares_fattree}, fat tree~\cite{leiserson_fattree},
Dragonfly+~\cite{dragonflyplus}, and rail-optimized fat
tree~\cite{railonly}. Every GPU has 18 links at 200\,Gb/s in each direction,
and each topology specifies its virtual-channel buffering. The pool choices
include low-power double-data-rate 5X (LPDDR5X), Non-Volatile Memory Express
(NVMe) SSD, and high-bandwidth flash (HBF). Pool size varies from 1 to
46 memory nodes, pooled data includes the KV cache, expert weights, or both,
and attachment is direct or through leaf switches. These choices define 64
pool structures on one NVL-style configuration and the rail-optimized fat
tree. Together with the 23 no-pool configurations, they yield 151 design
points. For the data placement, we store as much of the model
weights and KV cache as possible in HBM. The pool stores the overflow KV and weights,
while the remaining overflow uses host DRAM and then SSD when the pool is still insufficient.
Table~\ref{tab:case1-memory} shows the capacity, bandwidth, and access
latency of each memory tier.

\begin{table}[t]
\caption{Memory profiles used in Case Study I. Attachment and fabric bandwidth
can further limit the listed read bandwidth.}
\label{tab:case1-memory}
\scriptsize
\centering
\setlength{\tabcolsep}{2.5pt}
\begin{tabular}{@{}lccc@{}}
\toprule
Tier & Capacity & Read bandwidth & Fixed latency \\
\midrule
H200 HBM~\cite{nvidia_h200} & 141\,GB/GPU & 4.8\,TB/s & Not modeled \\
Host DRAM~\cite{nvidia_dgxh200,nvidia_h200,gpu_paged_memory} & 256\,GB/GPU & 64\,GB/s & 2\,$\mu$s \\
Host SSD~\cite{micron7450} & 3.84\,TB/GPU & 6.8\,GB/s & 80\,$\mu$s \\
LPDDR5X pool~\cite{nvidia_gh200_benchmark,werner_c2c} & 480\,GB/node & 335\,GB/s & 1\,$\mu$s \\
NVMe pool~\cite{micron9550} & 30.72\,TB/node & 14\,GB/s & 60\,$\mu$s \\
HBF pool~\cite{sandisk_hbf,h3_hbf} & 1.024\,TB/node & 3.2\,TB/s & 20\,$\mu$s \\
\bottomrule
\end{tabular}
\end{table}

\subsubsection{Results}
\label{sec:case-1-result}

\begin{figure}[t]
\centering
\includegraphics[width=\linewidth]{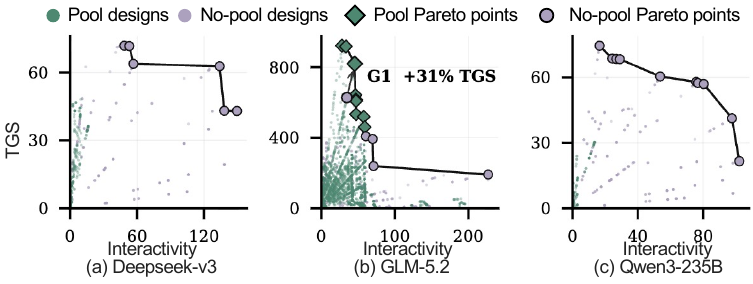}
\caption{Memory-pool and no-memory-pool serving frontiers at a one-million-token
context.}
\vspace{-15pt}
\label{fig:pool-pareto}
\end{figure}

As shown in Figure~\ref{fig:pool-pareto}, memory pools push the TGS-interactivity frontier forward for GLM-5.2, while all the memory-pool points on the frontier use HBF. Taking Point G1 in Figure~\ref{fig:pool-pareto}(b) as an example, it uses all 128
GPUs for both prefill and decode ($P/D=128/128$), forming eight replicas with
tensor-parallel degree 16.
It attaches a 46-node HBF KV pool to a rail-optimized fat tree. The memory pool
increases interactivity and TGS
by 1.31$\times$.

\begin{figure}[t]
\centering
\includegraphics[width=\linewidth,pagebox=cropbox]{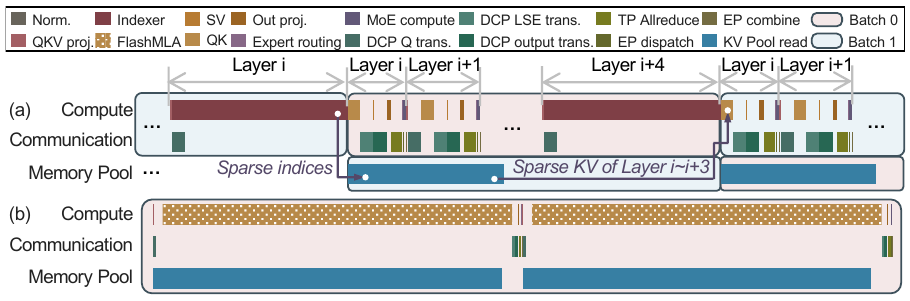}
\caption{\revised{Timelines for (a) sparse and (b) dense attentions.}}
\label{fig:timeline}
\end{figure}

The results show that the memory pools mainly benefit for sparse KV access.  GLM-5.2 uses sparse attention in 75 of its 78 layers. At a one-million-token
context, each sparse layer reads only the selected 2,048 tokens and its index summaries, corresponding to 0.54\% of the sparse-layer KV. The memory pool therefore expands KV capacity without placing the entire context on the fabric for every decode step. 
\revised{As shown in Figure~\ref{fig:timeline}(a), GLM-5.2 makes four layers share a common selection of sparse KV indices. So, the KV pool read can be hidden by four layers' execution of another batch. However, for dense attention in Figure~\ref{fig:timeline}(b), KV should be loaded in every FlashMLA attention layer, making GPUs waiting for pool read.}

\revised{As introduced in Section~\ref{sec:system}, we use \emph{ProtoBridge} to evaluate the top 15$\%$ interactivity and TGS designs. In this case study, using 15$\%$ threshold covers all the real pareto points. The total runtime is 58 minutes, with 51 minutes of surrogate evaluation and 7 minutes of packet-level refinement.}

\begin{figure}[t]
\centering
\includegraphics[width=\linewidth,pagebox=cropbox]{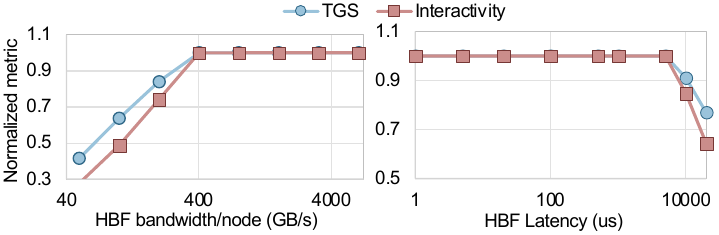}
\caption{G1 sensitivity to HBF read bandwidth and fixed latency. Values are
normalized to 3.2\,TB/s per node and 20\,$\mu$s.}
\label{fig:pool-sensitivity}
\end{figure}

\subsubsection{Insights}
\label{sec:case-1-insight}

\revised{First, memory pools mainly improve TGS for sparse attention.}
Memory pool provides additional capacity to increase batch size and allows the model to be deployed on a smaller cluster, but can hardly reduce every user's decode time in LLM serving. At G1 in Figure~\ref{fig:pool-pareto}, the pool
provides enough KV capacity for eight replicas with tensor-parallel degree 16
at batch 64, providing  much faster memory accesses than host DRAM and SSDs.  The additional
replicas and higher batch sizes raise TGS, while HBF latency and bandwidth prevent remote KV reads
from materially increasing TPOT.
Our findings are consistent with Mooncake~\cite{mooncake}, which independently demonstrates that disaggregated KV-cache storage can improve long-context serving throughput when its capacity and transfer bandwidth are sufficient.


\revised{
Second, scale-up memory pools can trade access latency for capacity and fabric scalability. The serving pipeline in Figure~\ref{fig:timeline}(a) tolerates additional fixed latency by overlapping remote KV reads with other decode operations, whereas insufficient bandwidth extends every large KV transfer and reduces both interactivity and throughput. As shown in Figure~\ref{fig:pool-sensitivity}, TGS and interactivity are less sensitive to the latency near G1. Memory-pool architectures can therefore use additional switch stages, longer-reach links, or higher-capacity media even when they increase fixed access latency. But they should provision enough parallel channels and memory-bank bandwidth to preserve sustained transfer bandwidth.
}


\subsection{Case Study II: Full-Optical CPO Fabrics}
\label{sec:case-cpo}

\subsubsection{Setup}

Electrical link loss rises with signaling rate and distance, limiting the
reach and energy efficiency of high-speed copper links. Optical transmission
has lower reach-dependent loss, while CPO shortens the electrical path between
the GPU or switch and the optical engine. Prior studies identify CPO as a promising way to achieve higher bandwidth density and lower energy per
bit~\cite{cpo_status,beyond_cpo,teraphy_vlsi,sipac_photonics,fowlp_cpo}.
Recent implementations report 32 channels at 200\,Gb/s with twice the
bandwidth density and 30\% lower power per bit than an electrical link~\cite{marvell_sipho200,intel_oci,
broadcom_bailly}.

We use optical-electrical (O/E) hybrid interconnection as the baseline, which selects the link medium by distance. It uses
passive copper for connections up to 2\,m, active electrical cable for
connections from 2\,m to 5\,m, and optical fiber beyond
5\,m~\cite{nvidia_linkx_reach}. We compare this reach-gated baseline with
full-optical CPO interconnections to determine whether their higher channel
count and data rate improve serving performance after accounting for optical
conversion latency and link-error handling.

\textbf{Design space.}
We use the same LLMs, output length, and DeepEP modes
as Section~\ref{sec:case-memory-pool}. Case Study II evaluates 1,024
GB200 GPUs, each with 186\,GB of HBM, 8\,TB/s of HBM bandwidth, and
2.5 peta floating-point operations per second of dense bfloat16 (BF16)
compute~\cite{nvl72}. Every input contains 1M tokens, and per-replica batch sizes are from 1 to 1024.

The O/E interconnection has 18 channels at 400\,Gb/s and admits 39 topology
configurations. The single-switch NVL-style fabric from
Section~\ref{sec:case-memory-pool} cannot span 1,024 GPUs, so we exclude it. We
add a routed 2D-full-mesh Clos~\cite{cloudmatrix384}, as we are going to explore CPO with more channels per GPU. CPO uses 32, 64, or 128 channels at 200, 400,
or 800\,Gb/s, and optical conversion adds 10\,ns at
each conversion interface~\cite{marvell_cpo,ayar_teraphy}. Combining each channel count with the three rates gives 492 CPO
interconnect designs. Together with the 39 O/E designs, the interconnect space
contains 531 points.
The pre-FEC BER ranges from $10^{-9}$ to $10^{-3}$~\cite{teraphy_vlsi,ieee_200g_pmd_fec}. For error handling, FEC is disabled or
uses RS$(528,514,7)$, RS$(544,514,15)$, or
RS$(576,514,31)$~\cite{rs_codes,kp4_fec}. Retransmission uses
GBN or SACK with a buffer of 0 to 8\,MB.  We evaluate every design point
at the six BER values. 
For each LLM, we form the Pareto frontier of the 18$\times$400-Gb/s O/E
baseline. We choose the point with the largest product
of the TGS and interactivity as the reference for the following analysis. This
choice balances the two metrics rather than favoring either endpoint.

\subsubsection{Results}

\begin{figure*}[!t]
\centering
\includegraphics[width=0.99\textwidth]{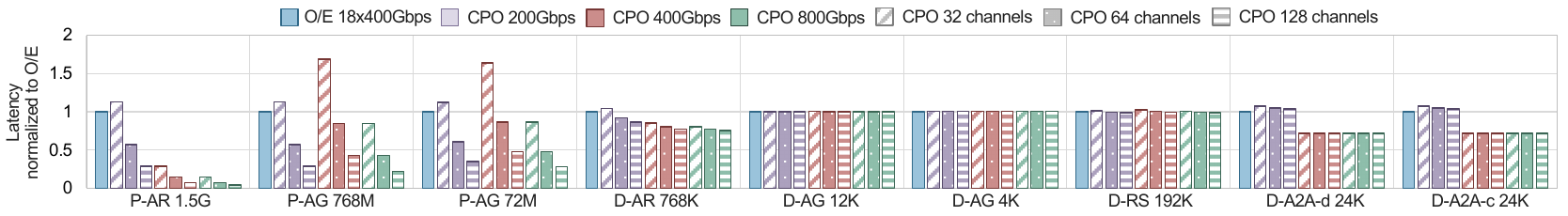}
\caption{Operation latency normalized to the corresponding O/E interconnection.
Lower is better; O/E baseline equals 1.}
\vspace{-15pt}
\label{fig:cpo-operation}
\end{figure*}

Figure~\ref{fig:cpo-operation}  uses GLM-5.2's operators as an example, isolating nine communication operations with various data sizes from
each design, including prefill AllReduce (P-AR) and AllGather (P-AG), decode AllReduce (D-AR), AllGather (D-AG), and AlltoAll dispatch (D-A2A-d) and combine (D-A2A-c). We simulate each operation on the O/E
interconnection and all nine CPO configurations.
The CPO runs use pre-FEC BER
$10^{-9}$, SACK, and a 4\,MB retransmission buffer.
Relative to O/E, 128 CPO channels at 800\,Gb/s make the three prefill
operations 
3.55$\times$$\sim$24.8$\times$ faster for GLM-5.2, while speedups for the decode
operations are 0.99$\times$$\sim$1.40$\times$.
The results show that for prefill communications with large data sizes, high bandwidth CPO greatly reduces communication latency. \revised{For decode communications under our configurations, increasing bandwidth hardly brings benefits as they are latency-dominated}; instead, the conversion latency and error handling overhead of CPO may increase the communication latency. However, for decode AlltoAll, increasing channel number provides a 39\% improvement. This is because increasing channel number gives us a larger topology design space. 64- and 128-channel designs use 2D-full-mesh topology, compared to leaf-spine topology in the 32-channel design, the network contention is greatly reduced for AlltoAll communications.

\begin{table}[!t]
\centering
\small
\setlength{\tabcolsep}{4pt}
\caption{Serving latency for O/E and 128$\times$800-Gb/s CPO at BER
$10^{-9}$. Each entry shows O/E $\rightarrow$ CPO.}
\label{tab:cpo-serving-latency}
\begin{tabular}{lcc}
\toprule
Model & TTFT (s) & TPOT (ms) \\
\midrule
DeepSeek-V3 & 63.49 $\rightarrow$ 63.22 & 24.75 $\rightarrow$ 24.02 \\
GLM-5.2 & 3.834 $\rightarrow$ 3.543 & 9.252 $\rightarrow$ 8.386 \\
Qwen3-235B & 39.10 $\rightarrow$ 38.85 & 12.79 $\rightarrow$ 13.11 \\
\bottomrule
\end{tabular}
\end{table}

Table~\ref{tab:cpo-serving-latency} shows how CPO changes the TTFT and TPOT for three models. Using CPO with 128$\times$800 Gb/s per GPU reduces TTFT by 0.42\%$\sim$7.58\%, reduces TPOT by 2.93\% for DeepSeek-V3 and 9.36\% for GLM-5.2, but increases TPOT by 2.50\% for Qwen3-235B. Although CPO reduces much prefill communication overhead, the TTFT only decreases slightly as prefill is compute-bound.

\begin{figure}[!t]
\centering
\includegraphics[width=\columnwidth]{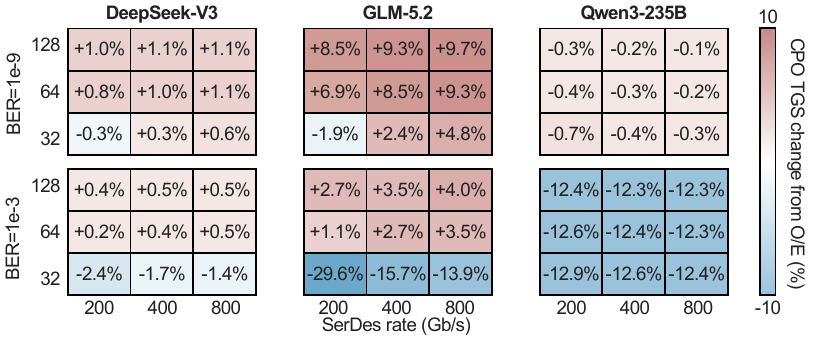}
\caption{Effect of pre-FEC BER on CPO TGS relative to O/E at 1,024 GPUs.}
\vspace{-15pt}
\label{fig:cpo-summary}
\end{figure}

Figure~\ref{fig:cpo-summary} compares CPO at pre-FEC BER $10^{-9}$ and
$10^{-3}$. 
At BER $10^{-3}$, the best
128$\times$800-Gb/s design raises TGS by 0.54\% for DeepSeek-V3 and 3.98\%
for GLM-5.2, but reduces Qwen3-235B TGS by 12.30\%. Even with the strongest FEC RS$(576,514,31)$ among our design space and the fastest retransmission strategy SACK, a high BER of $10^{-3}$ nearly destroys all improvements that CPO brings.
\revised{We also tested the ranking robustness to timing uncertainty. Under 10\% communication timing noise, the CPO ranking is unchanged and the TTFT and TPOT changes are below 5.3\%.}

\subsubsection{Insights}
\label{sec:case-2-insight}
First, operation size determines whether CPO bandwidth is useful.
Large prefill operations amortize protocol startup and optical conversion latency, so
their latency falls substantially as per-GPU interconnection bandwidth rises. However, most decode
operations are too small to do so. \revised{Under our configuration, Qwen3-235B's 512-B expert All-to-All is slower on CPO, making its TPOT
increase.}  Furthermore, higher BER of CPO introduces more FEC overhead and retransmission latency, making the TPOT even worse. \revised{In short, under our design space, CPO's high bandwidth can hardly benefit LLM decode performance unless the high BER problem get solved in the future.}

\revised{
Second, use the optical I/O budget to increase connectivity before increasing per-channel rate. At thousand-GPU scale, additional channels can change the feasible topology and remove shared-path contention, while a higher rate on the same number of channels preserves the original bottlenecks. CPO designs should therefore prioritize channel count and topology construction for communication-intensive MoE traffic, and increase SerDes rate only after sufficient path diversity is available. UB-mesh~\cite{cloudmatrix384} provides an independent support of our finding, by designing an ND-full-mesh topology with optical interconnections.}

\section{Related Work and Discussion}
\label{sec:related}


\textbf{System simulators.}
Chakra standardizes distributed machine-learning operations and their
dependencies~\cite{chakra}. ASTRA-sim consumes such workloads and models
collectives over analytical or detailed network
backends~\cite{astra_sim,astra_sim2,astra_sim3}. A recent extension to Chakra
also represents collective algorithms as point-to-point transfers and
reductions~\cite{collective_representation}. SimAI increases training-system
fidelity by integrating framework, kernel, collective, and network
models~\cite{simai}, while Phantora executes unmodified training frameworks and
intercepts their compute and communication operations~\cite{phantora}.
LLMServingSim, Vidur, and Frontier instead model serving schedules,
disaggregation, and runtime state~\cite{llmservingsim,llmservingsim2,vidur,frontier}.
These systems describe when communication operations or collective transfers
execute. \emph{ProtoBridge} continues from
those operations to protocol packetization, completion dependencies, and the
packet mechanisms selected for a scale-up design.

\textbf{Detailed network simulation.}
ns-3, BookSim, Garnet, and HTSim resolve packet timing, queues, routing, and
switch behavior~\cite{ns3,booksim,garnet,htsim}. \revised{However, BookSim, Garnet, and HTSim are cycle-accurate NoC simulators. They are unsuitable for scale-up simulation because they lack protocol-level mechanisms such as flow control, and their cycle-accurate time advancement is slower than ns-3's discrete-event approach.} CNSim adds cycle-accurate,
packet-parallel simulation for large chiplet networks, and Arcadia targets
cluster-scale network simulation for AI
workloads~\cite{cnsim,arcadia}. SimBricks composes
independent simulators~\cite{simbricks}. DICE models physical-layer coding,
channel errors, retransmission, and flow control along an inter-chiplet
datapath~\cite{dice}. These tools execute traffic supplied by a workload or
protocol implementation. Adding a new scale-up protocol still requires code
that generates its traffic and maps packet outcomes to operation completion.
\emph{ProtoBridge} factors those choices into its two-layer configuration. 
\revised{Wormhole~\cite{wormhole} accelerates training simulation by caching recurring contention patterns. In LLM serving, decode operations are short and non-repetitive, limiting Wormhole's caching benefit. 
}

\textbf{Analytical performance models.}
FlexFlow and Proteus estimate distributed execution while searching
parallelization strategies~\cite{flexflow,proteus}. Calculon and LLMCompass use
parameterized analytical models to explore LLM system and hardware
designs~\cite{calculon,llmcompass}. Such models make broad exploration
tractable by representing communication with aggregate latency and bandwidth
parameters. Our surrogate instead retains separate terms for
the protocol transfer and for delays introduced by flow control, contention,
coding, and retransmission. Section~\ref{sec:evaluation} compares the resulting
prediction and design-ranking accuracy.

\revised{
\textbf{Extensibility and future work.} \emph{ProtoBridge} currently targets LLM serving over scale-up fabrics. Supporting training and HPC workloads requires workload-specific schedules in OTP while retaining the existing PEX mechanisms. Extending \emph{ProtoBridge} to scale-out protocols such as RDMA and TCP requires additional engineering effort and remains future work.}



\section{Conclusion}
\label{sec:conclusion}

\revised{In this paper, we present \emph{ProtoBridge}, an ns-3-based LLM-to-packet simulator that represents protocol-defined packet exchanges through OTP and packet-execution mechanisms through PEX.} \emph{ProtoBridge} achieves average errors of 17.8$\%$ on H200 NVLink4 and 10.3$\%$ on C500 MetaXLink. We also propose the protocol-defined surrogate that enables efficient exploration followed by packet-level validation. The memory-pool and CPO case studies demonstrate how protocol and interconnect choices affect emerging LLM superpod designs.


\section*{Acknowledgments}
This work was supported by the National Key R\&D Program of China (2025YFB4507300), the National Natural Science Foundation of China (62504139, 62325405), Shenzhen Science and Technology Program (AI2026018, AI2026015), Beijing Natural Science Foundation (L257010, L242018), Beijing National Research Center for Information Science, Technology (No. BNR2024TD03001), Beijing Municipal Science and Technology Project (Z241100004224025), and State Key Laboratory of Space Network and Communications.
This research was partially supported by ACCESS – AI Chip Center for Emerging Smart Systems, sponsored by InnoHK funding, Hong Kong SAR (HKSAR) and Research Grants Council of HKSAR (16213824).

\bibliographystyle{IEEEtran}
\bibliography{main}


\end{document}